# BRANS-DICKE CORRECTIONS TO THE GRAVITATIONAL SAGNAC EFFECT

K.K.Nandi[1,2], P.M.Alsing[3], J.C.Evans[4], and T.B.Nayak[1]

1) Department of Mathematics, University of North Bengal, Darjeeling (W.B.) 734 430, India. E-mail: kamalnandi@hotmail.com
2) Senior Research Associate, Inter-University Center for Astronomy and Astrophysics (IUCAA), Ganeshkhind, Pune (M.S.) 411 007, India.
3) Albuquerque High Performance Computing Center, 1601 Central NE, University of New Mexico, Albuquerque, NM 87131, U.S.A. E-mail: alsing@ahpcc.unm.edu
4) Department of Physics, University of Puget Sound, Tacoma, WA 98416, U.S.A. E-mail: jcevans@ups.edu

## Abstract

The *exact* formulation for the effect of the Brans-Dicke scalar field on the gravitational corrections to the Sagnac delay in the Jordan and Einstein frames is presented for the first time. The results completely agree with the known PPN factors in the weak field region. The calculations also reveal how the Brans-Dicke coupling parameter $\varpi$ appears in various correction terms for different types of source/observer orbits. A first order correction of roughly $2.83 \times 10^{-1}$ fringe shift for visible light is introduced by the gravity-scalar field combination for Earth bound equatorial orbits. It is also demonstrated that the final predictions in the two frames do not differ. The effect of the scalar field on the geodetic and Lense-Thirring precession of a spherical gyroscope in circular polar orbit around the Earth is also computed with an eye towards the Stanford Gravity Probe-B experiment currently in progress. The feasibility of optical and matter-wave interferometric measurements is discussed briefly.

# 1. INTRODUCTION

Ever since its discovery, Sagnac effect [1] has played a very important role in the understanding and development of fundamental physics. For a recent review, see the works of Stedman [2]. The effect stems from the basic physical fact that the round-trip time of light around a closed contour, when its source is fixed on a turntable, depends upon the angular velocity, say $\Omega$, of the turntable. Furthermore, this round-trip time is different for light co-rotating and counter-rotating with the turntable. Using Special Theory of Relativity (STR), and assuming $\Omega r \ll c$, one obtains the proper time difference $\delta\tau_S$ when the two beams meet again at the starting point as [3]:

$$\delta\tau_S \cong \frac{4\Omega}{c^2} S, \qquad (1)$$

where c is the vacuum speed of light, S ($=\pi r^2$) is the projected area of the contour perpendicular to the axis of rotation. Note that the expression (1) represents a lack of simultaneity as recorded by a single rotating clock (from where the beams depart and reunite). It is thus a real physical effect in the sense that it does not involve any arbitrary synchronization convention that is required between two distant clocks [3,4]. Moreover, the effect is universal as it manifests not only for light rays but also for all kinds of waves including matter waves [5-11].

The formula (1) has been tested to a good accuracy and the remarkable degree of precision attained lately by the advent of ring laser interferometry raises the hope that the measurements of higher order corrections to this effect might be possible in the near future [2]. Motivated by this prospect, Tartaglia [12], in a recent interesting paper, has considered the Einsteinian General Relativistic (EGR) effects on the proper delay time when the source/receiver orbits a massive rotating body (a "massive turntable", as it were). The author considered the Kerr metric for a rotating body and obtained the EGR corrections to the Sagnac effect in the cases when the light source/receiver executes equatorial, polar and geodesic circular motions.

On the other hand, there is a recent surge of interest in the non-Einsteinian theories of gravity such as the celebrated Brans-Dicke (BD) theory [13] or other scalar tensor theories. The motivation comes from the fact that the occurrence of scalar fields coupled to gravity seems inevitable in superstring theories [14], higher order theories [15] as well as in the extended [16] and hyperextended [17] inflationary theories of the Early Universe. Moreover, scalar tensor theories provide a rich arena for investigations into wormhole physics [18-23]. One also recalls that the standard solar system tests of gravity were calculated in the BD theory that displayed the effect of the scalar field on those tests. Current experimental estimates place the BD coupling parameter $\varpi \geq 500$. In the same spirit, it seems quite desirable that the effect of the scalar field on the corrections to the Sagnac effect, geodetic and Lense-Thirring precesion be also calculated using a Kerr-like solution of the BD theory. This precisely is the aim of the present paper, and we follow exactly the same procedure as in Ref.[12] for the Sagnac part.

In dealing with scalar-tensor theories in general and BD theory in particular, one envisages two types of variables delineating two types of frames, viz., the Jordan and Einstein frames which are connected by the scalar field. In Sec.2, we discuss the rotating solutions in the two frames. Sections 3 and 4 derive, respectively, the exact and approximate expressions for the proper time delay $\delta\tau$ in the case of the equatorial trajectory of the source/observer. The polar and geodesic trajectories are considered in Sections 5 and 6 respectively. In Sec.7, the relevant corrections in the Einstein frame are considered. Sec.8 contains a broad discussion which is divided into various subsections containing numerical estimates for the Sagnac delay in STR and BD theory for Earth-bound experiments, a comparison with the usual PPN factors as well as the possibility of using optical and matter-wave interferometers to measure the correction factors. In Sec.9, we calculate the geodetic and Lense-Thirring precession in the weak field limit of the Kerr-like BD metric for a satellite in a circular polar orbit about the Earth. We end with a summary of our results in Sec.10.

## 2. ROTATING SOLUTIONS IN THE JORDAN AND EINSTEIN FRAMES

Let us first define what are meant by the Jordan and Einstein frames [15,20]. The pair of variables ($g_{\mu\nu}$, scalar $\phi$) defined originally in the BD action constitute what is called a Jordan frame. Consider now the conformal rescaling

$$\widetilde{g}_{\mu\nu} = f(\phi) g_{\mu\nu}, \widetilde{\phi} = h(\phi) \tag{2}$$

such that, in the redefined action, $\widetilde{\phi}$ couples minimally to $\widetilde{g}_{\mu\nu}$ for some functions f(φ) and h(φ). Then the new pair ($\widetilde{g}_{\mu\nu}$, scalar $\widetilde{\phi}$) is said to constitute an Einstein frame. Sometimes, it is mathematically more preferable to use this latter frame for computation of experimental predictions. In the Jordan pair, the scalar field φ plays the role of a component of gravity in the sense that <φ>≈G$^{-1}$, where G is the Newtonian constant of gravity, signifying the Machian character of the BD theory. On the other hand, in the Einstein pair, the scalar $\widetilde{\phi}$ plays the role of some kind of matter source. These features will become evident from the field equations that follow. Throughout this paper, we take G=c=1 unless they are explicitly restored.

The matter-free Jordan frame BD action is given by:

$$S_J[g_{\mu\nu},\phi] = \frac{1}{16\pi}\int\left(\phi R - \frac{\varpi}{\phi}g^{\mu\nu}\phi_{,\mu}\phi_{,\nu}\right)\sqrt{-g}d^4x \tag{3}$$

where $\varpi$ = constant is a dimensionless coupling parameter. The resultant field equations are

$$\left(\phi^{;\rho}\right)_{,\rho} = 0, \tag{4}$$

$$R_{\mu\nu} - \frac{1}{2}g_{\mu\nu}R = \frac{\varpi}{\phi^2}\left[\phi_{,\mu}\phi_{,\nu} - \frac{1}{2}g_{\mu\nu}\phi_{,\sigma}\phi^{,\sigma}\right] + \frac{1}{\phi}\left[\phi_{;\mu;\nu} - g_{\mu\nu}\left(\phi^{;\sigma}\right)_{;\sigma}\right] \tag{5}$$

where the ; indicates covariant derivative with respect to $g_{\mu\nu}$. Following the procedure of Newman and Janis [24], a two-parameter rotating solution of the above field equations has indeed been found by Krori and Bhattacharjee [25] from the static BD solution. They called it a Kerr-like solution but we choose to call it the KB solution in what follows. In order to see how the different arbitrary constants are related, it is necessary to display the static BD solution which, in "isotropic" coordinates, $(t,\bar{\rho},\theta,\varphi)$ is:

$$ds^2 = \left[\frac{1-\frac{r_0}{2\bar{\rho}}}{1+\frac{r_0}{2\bar{\rho}}}\right]^{\frac{2}{\lambda}} dt^2 - \left(1+\frac{r_0}{2\bar{\rho}}\right)^4\left[\frac{1-\frac{r_0}{2\bar{\rho}}}{1+\frac{r_0}{2\bar{\rho}}}\right]^{\frac{2(\lambda-C-1)}{\lambda}}\left[d\bar{\rho}^2 + \bar{\rho}^2 d\theta^2 + \bar{\rho}^2 \sin^2\theta\, d\varphi^2\right] \tag{6}$$

$$\phi = \phi_0\left[\frac{1-\frac{r_0}{2\bar{\rho}}}{1+\frac{r_0}{2\bar{\rho}}}\right]^{\frac{C}{\lambda}}, \tag{7}$$

where $\lambda$, $C$, $\phi_0$, $r_0$ are constants, and the first two relate to $\varpi$ as

$$\lambda^2 \equiv (C+1)^2 - C\left(1 - \frac{\varpi C}{2}\right) \tag{8}$$

The KB solution generated from the above is given by

$$ds^2 = g_{\mu\nu}dx^\mu dx^\nu$$

$$= \left(1 - \frac{2r_0 r}{\rho}\right)^\eta (dt - \omega d\varphi)^2 - \left(1 - \frac{2r_0 r}{\rho}\right)^\xi \rho\left(\frac{dr^2}{\Delta} + d\theta^2 + \sin^2\theta d\varphi^2\right)$$

$$+ 2\left(1 - \frac{2r_0 r}{\rho}\right)^\sigma \omega(dt - \omega d\varphi)d\varphi, \tag{9}$$

$$\phi = \phi_0\left(1 - \frac{2r_0 r}{\rho}\right)^{-\sigma}, \sigma = \frac{\xi + \eta - 1}{2} = -\frac{C}{2\lambda}. \tag{10}$$

$$\omega = a\sin^2\theta, \ \rho = r^2 + a^2\cos^2\theta, \ \Delta = r^2 + a^2 - 2r_0 r, \ r = \bar{\rho}(1 + r_0/2\bar{\rho})^2. \tag{11}$$

The solutions (9)-(11) represent the exterior metric due to a massive body rotating with respect to the fixed stars, the scalar field being given by Eq.(10). As one can see, the presence of the coupling parameter $\varpi$ in the solution is manifested through the expressions (8) and (10). For $\xi=0$, $\sigma=0$, $\eta=1$, one recovers the usual Kerr metric in Boyer-Lindquist coordinates. Here $r_0 = GM/c^2$, M is the mass of the source and a is the ratio between the total angular momentum J and the mass M, that is, a=J/M.

The Einstein frame action is obtained from the BD action (3) by means of a particular conformal transformation, called the Dicke transformations, given by

$$\tilde{g}_{\mu\nu} = \frac{1}{16\pi}\phi g_{\mu\nu} \tag{12}$$

$$d\tilde{\phi} = \left(\frac{\varpi + \frac{3}{2}}{\alpha}\right)^{\frac{1}{2}} \frac{d\phi}{\phi}, \tag{13}$$

where $\alpha$ is an arbitrary constant. The action then is

$$S_E[\widetilde{g}_{\mu\nu}, \widetilde{\phi}] = \int [\widetilde{R} - \alpha \widetilde{g}^{\mu\nu} \widetilde{\phi}_{,\mu} \widetilde{\phi}_{,\nu}] \sqrt{-\widetilde{g}} d^4 x. \tag{14}$$

The resulting field equations are

$$\widetilde{R}_{\mu\nu} - \frac{1}{2} \widetilde{g}_{\mu\nu} \widetilde{R} = \alpha \left[ \widetilde{\phi}_{,\mu} \widetilde{\phi}_{,\nu} - \frac{1}{2} \widetilde{g}_{\mu\nu} \widetilde{\phi}_{,\sigma} \widetilde{\phi}^{,\sigma} \right] \tag{15}$$

$$\left( \widetilde{\phi}^{;\rho} \right)_{,\rho} = 0 \tag{16}$$

The KB solutions of the above Einstein-minimally coupled equations (15), (16) can be explicitly written out as:

$$ds^2 = \widetilde{g}_{\mu\nu} dx^\mu dx^\nu$$
$$= \left(1 - \frac{2r_0 r}{\rho}\right)^{\eta-\sigma} (dt - \omega d\varphi)^2 - \left(1 - \frac{2r_0 r}{\rho}\right)^{\xi-\sigma} \rho \left(\frac{dr^2}{\Delta} + d\theta^2 + \sin^2\theta d\varphi^2\right)$$
$$+ 2\omega(dt - \omega d\varphi)d\varphi, \tag{17}$$

$$\widetilde{\phi} = -\left[\frac{\varpi + \frac{3}{2}}{\alpha}\right]^{\frac{1}{2}} \sigma \ln\left(1 - \frac{2r_0 r}{\rho}\right), \tag{18}$$

$$\omega = a \sin^2 \theta, \rho = r^2 + a^2 \cos^2 \theta, \Delta = r^2 + a^2 - 2r_0 r. \tag{11}$$

Here also, for a=0, the solutions (17), (18) go over to Buchdahl solutions [20,26] in "standard" coordinates under a suitable radial transformation defined below.

The vacuum KB solution (9) resembling the Kerr metric is defined for the radial coordinate r in the range $r_0 + (r_0^2 - a^2 \cos^2 \theta)^{\frac{1}{2}} < r < \infty$ which translates in "standard" radial coordinate $\overline{R}$ into the range $0 < \overline{R} < \infty$ where $\overline{R}$ is defined by

$$\overline{R}^2 = \rho \left(1 - \frac{2r_0 r}{\rho}\right)^\xi. \tag{19}$$

The solution does exhibit a curvature singularity at the origin $\overline{R} = 0$ which is not clothed by an event horizon and hence is naked. In fact, the singularity has the topology of a point as the area of the equipotential surfaces and proper lengths of closed curves on these surfaces all reduce to zero size as $\bar{R} \to 0.$ The coupling between gravity and a massless scalar field renders the event horizon to collapse to a point and one has gravitation without black holes [27]. At any rate, we are interested only in the effects due to a normal, uncollapsed rotating star coupled to a scalar field. Hence, the Penrose Conjecture of Cosmic Censorship (preventing the occurrence of naked singularities), for which a precise formulation is yet unavailable, should not concern us here. Indeed, we will see that the PPN calculations precisely agree with those following from the KB metrics in both Jordan and Einstein frames.

## 3. EQUATORIAL TRAJECTORY

Consider that the source/receiver of two oppositely directed light beams is moving around the gravitating body, along a circumference at a radius r=R=constant ($R > r_0 + (r_0^2 - a^2 \cos^2 \theta)^{\frac{1}{2}}$) on the equatorial plane $\theta = \pi/2$. Suitably placed mirrors send back to their origin both beams after a circular trip about the central body. Let us further assume that the source/receiver is moving with uniform orbital angular speed $\omega_0$ with respect to distant stars such that the rotation angle is

$$\varphi_0 = \omega_0 t. \tag{20}$$

Under these conditions, the KB metric (9) reduces to:

$$ds^2 = [\chi \omega_0^2 + 2a(P^\sigma - P^\eta)\omega_0 + P^\eta]dt^2, \chi \equiv P^\eta a^2 - P^\xi R^2 - 2P^\sigma a^2. \tag{21}$$

$$P = \left(1 - \frac{2r_0}{R}\right). \tag{22}$$

The trajectory of a light ray is given by $ds^2=0$ which immediately gives

$$0 = \chi \omega^2 + 2a(P^\sigma - P^\eta)\omega + P^\eta \equiv \chi(\omega - \Omega_+)(\omega - \Omega_-) \tag{23}$$

where $\omega$ is the orbital angular speed of photons. The two roots $\Omega_\pm$ satisfy the following equations

$$\Omega_+ + \Omega_- = -\frac{2a(P^\sigma - P^\eta)}{\chi}, \Omega_+ \Omega_- = \frac{P^\eta}{\chi}. \tag{24}$$

The rotation angles for light are then

$$\varphi_\pm = \Omega_\pm t. \tag{25}$$

Eliminating t between eqs.(20) and (25), we get

$$\varphi_\pm = \frac{\Omega_\pm}{\omega_0} \varphi_0. \tag{26}$$

The first intersection of the world lines of the two light rays with the world line of the orbiting observer after emission at time t=0 occurs when

$$\varphi_+ = \varphi_0 + 2\pi, \varphi_- = \varphi_0 - 2\pi, \text{or}, \frac{\Omega_\pm}{\omega} \varphi_0 = \varphi_0 \pm 2\pi \tag{27}$$

where + refers to corotating and - refers to counterrotating beams. Solving for $\varphi_0$, we get

$$\varphi_{0\pm} = \pm \frac{2\pi \omega_0}{\Omega_\pm - \omega_0}. \tag{28}$$

The proper time of the rotating observer is deduced from Eq.(21) as

$$d\tau = \sqrt{[\chi \omega_0^2 + 2a(P^\sigma - P^\eta)\omega_0 + P^\eta]} \frac{d\varphi_0}{\omega_0} \tag{29}$$

Therefore, integrating between $\varphi_{0+}$ and $\varphi_{0-}$, we obtain the Sagnac delay

$$d\tau = \sqrt{[\chi \omega_0^2 + 2a(P^\sigma - P^\eta)\omega_0 + P^\eta]} \frac{\varphi_{0+} - \varphi_{0-}}{\omega_0} \tag{30}$$

From Eq.(28), we have,

$$\varphi_{0+} - \varphi_{0-} = 2\pi \omega_0 \left[ \frac{\Omega_+ + \Omega_- - 2\omega_0}{(\Omega_+ - \omega_0)(\Omega_- - \omega_0)} \right] \tag{31}$$

Using this expression in Eq.(30), we find

$$\delta\tau = (2\pi)\frac{\chi[(\Omega_+ + \Omega_-) - 2\omega_0]}{\sqrt{\chi\omega_0^2 + 2a(P^\sigma - P^\eta)\omega_0 + P^\eta}}. \tag{32}$$

We see that the delay $\delta\tau$ is zero if the angular speed of the orbiting observer is

$$\omega_0 \equiv \omega_n = \frac{\Omega_+ + \Omega_-}{2} = \frac{a(P^\eta - P^\sigma)}{\chi} = \frac{a(P^\eta - P^\sigma)}{P^\eta a^2 - P^\xi R^2 - 2P^\sigma a^2}. \tag{33}$$

provided a≠0. In the usual Kerr case, the above reduces to ($r_0 = M$):

$$\omega_n = \frac{2aM}{R^3 + a^2 R + 2Ma^2} \tag{34}$$

which is exactly the same as the one obtained by Tartaglia [12]. The observers having the angular speed $\omega_n$ are locally nonrotating and may be imagined to be equivalent to the static observers in the Schwarzschild geometry for whom no Sagnac effect exists. On the other hand, if the observers keep fixed positions with regard to distant stars so that $\omega_0$=0, then the Sagnac delay becomes

$$\delta\tau_0 = \delta\tau(\omega_0 = 0) = (4\pi a)\frac{(P^\eta - P^\sigma)}{\sqrt{P^\eta}}. \tag{35}$$

In the usual Kerr case, one obtains from the above

$$\delta\tau_0 = \frac{8\pi aM}{R\sqrt{1 - \frac{2M}{R}}} = \frac{8\pi J}{R\sqrt{1 - \frac{2M}{R}}} = \frac{8\pi I_0 \Omega_0}{R\sqrt{1 - \frac{2M}{R}}} \tag{36}$$

in which we have used the expression for the moment of inertia $I_0$ given by J=aM= $I_0 \Omega_0$ where $\Omega_0$ is the angular speed of the rotating source, assumed to be solid and spherical with uniform density. The expression (36) again is the same as in Ref.[12].

To the order in $1/R^2$, we have, from Eq.(35),

$$\delta\tau_0 \cong \frac{8\pi a r_0}{R}(\sigma - \eta)\left[1 + \frac{r_0}{R}(1 - \sigma - \eta)\right]\left[1 - \frac{2r_0}{R}\right]^{-\frac{\eta}{2}}.$$

$$\cong \frac{8\pi a r_0}{R}(\sigma - \eta)\left[1 + \frac{r_0}{R}(1-\sigma)\right]. \tag{37}$$

This too coincides with the calculations in the Kerr case when appropriate values σ and η are chosen. However, the effect of the scalar field is manifest in the detemination of values for σ and η away from the Kerr values.

One may also reexpress the delay $\delta\tau_0$ in terms of the Lense-Thirring effect (see Section 9) given by (using $ar_0 = I\Omega_0$):

$$\omega_{LT} = \frac{I\Omega_0}{R^3} \tag{38}$$

and the result is

$$\delta\tau_0 \cong \frac{8\omega_{LT}}{R}(\sigma - \eta)(\pi R^3)\left[1 + \frac{r_0}{R}(1-\sigma)\right]. \tag{39}$$

If the observer is fixed on the equator, then $\omega_0 = \Omega_0$, then the delay $\delta\tau$ can also be expressed in terms of I, $r_0$ and $\Omega_0$:

$$\delta\tau = (4\pi\Omega_0)\frac{\frac{I}{r_0}(P^\eta - P^\sigma) - \chi}{\sqrt{\chi\Omega_0^2 + \frac{2I\Omega_0^2}{r_0}(P^\sigma - P^\eta) + P^\eta}} \tag{40}$$

where

$$\chi = \frac{I^2\Omega_0^2}{r_0^2}P^\eta - R^2 P^\xi - 2P^\sigma \frac{I^2\Omega_0^2}{r_0^2}. \tag{41}$$

All these reduce to the corresponding expressions in the Kerr case.

## 4. APPROXIMATIONS

For our convenience, let us adopt the following abbreviations

$$\zeta \equiv a/R, \quad \psi \equiv \omega_0 R, \quad \varepsilon \equiv r_0/R. \tag{42}$$

Since we shall be concerned mainly with Earth bound experiments, it is useful to have an idea of how small the quantities $\zeta$, $\psi$ and $\varepsilon$ are. For Earth, these are (Exact individual values of the pieces will be given later),

$$\zeta_\oplus = \frac{a_\oplus}{R_\oplus c} \sim 10^{-6}, \quad \psi_\oplus = \frac{\omega_0 R_\oplus}{c} \sim 10^{-7}, \quad \varepsilon_\oplus = \frac{GM_\oplus}{R_\oplus c^2} \sim 10^{-9}, \tag{43}$$

and for Sun, these are:

$$\zeta_\odot \sim \psi_\odot \sim \varepsilon_\odot \sim 10^{-6}. \tag{44}$$

Let us rewrite Eq.(32) as

$$\frac{\delta\tau}{4\pi R} = \frac{\psi(\chi/R^2) + \zeta(P^\sigma - P^\eta)}{\left[P^\eta + \psi^2(\chi/R^2) + 2\zeta\psi(P^\sigma - P^\eta)\right]^{1/2}}. \tag{45}$$

With the values displayed in eqs.(43) and (44) in mind, we use the expansions:

$$\frac{\chi}{R^2} = \zeta^2 P^\eta - P^\xi - 2\zeta^2 P^\sigma \cong -1 + 2\xi\varepsilon + 2\xi(\xi-1)\varepsilon^2 + \zeta^2 + O(\ )^3, \tag{46}$$

$$P^\sigma - P^\eta = \left[1 - \frac{2r_0}{R}\right]^\sigma - \left[1 - \frac{2r_0}{R}\right]^\eta \cong -2(\sigma - \eta)\varepsilon[1 + (1 - \sigma - \eta)\varepsilon] + O(\ )^3, \tag{47}$$

$$P^\eta = \left[1 - \frac{2r_0}{R}\right]^\eta \cong 1 - 2\eta\varepsilon + 2\eta(\eta - 1)\varepsilon^2 + O(\ )^3, \tag{48}$$

where $O(\ )^3$ stands for any cubic terms in the small quantities $\zeta$, $\psi$, $\varepsilon$. Using these expansions, we obtain the delay, denoting it by $\delta\tau_E$:

$$\begin{aligned}\frac{\delta\tau_E}{4\pi R} &\cong \psi + 2(\sigma - \eta)\varepsilon\zeta + (\eta - 2\xi)\varepsilon\psi \\ &\quad + \zeta^2\psi + 2(\eta - \sigma)(\sigma - 1)\varepsilon^2\psi \\ &\quad - \frac{1}{2}\left(4\eta\xi + 4\xi - 4\xi^2 - 2\eta - \eta^2\right)\varepsilon^2\psi + \frac{1}{2}\psi^3 + O(\ )^4. \end{aligned} \tag{49}$$

After cross multiplying and substituting in the definitions of small quantities in Eq.(49), we get:

$$\delta\tau_E \cong \delta\tau_S + \frac{8\pi r_0 a}{R}(\sigma - \eta) + 4\pi\omega_0 r_0 R(\eta - 2\xi)$$

$$+ 4\pi a^2 \omega_0 + \frac{8\pi r_0^2 a}{R}(\eta - \sigma)(\sigma - 1)$$

$$- 2\pi\omega_0^2 r_0^2 \left(4\eta\xi + 4\xi - 4\xi^2 - \eta^2 - 2\eta\right) + 2\pi\omega_0^3 R^4 + O(\ )^4. \tag{50}$$

The second term above represents the correction due to the moment of inertia I of the rotating source ($ar_0 = I\Omega_0$), the third term represents the correction due to the mass parameter $r_0$ and the remaining higher order terms represent variously combined effects of I, $r_0$ and $\Omega_0$. Most importantly, one can now visualize the effects of the scalar field through the factors $\eta$, $\sigma$ and $\xi$.

In the absence of a scalar field and for a homogeneous spherical object whose radius is $R_0$, one has:

$$I = \frac{8}{15}\pi\rho R_0^5 = \frac{2}{5} M R_0^2. \tag{51}$$

$\rho$ is the density (assumed to be uniform) of the object. Hence $a$ for the sphere is approximately

$$a \cong \frac{2}{5} R_0^2 \Omega_0. \tag{52}$$

## 5. POLAR (CIRCULAR) ORBITS

We shall now investigate the effect when the light rays move along a circular trajectory passing over the poles. In this case, too, we may take r=R =constant and φ=constant. Assuming uniform motion again, we take θ=$\omega_0$t. Then, we have, using dr=0, dφ=0, dθ=$\omega_0$dt and ds²=0, from the metric (9):

$$\frac{d\theta}{dt} = \pm \frac{(R^2 - 2r_0 R + a^2 - a^2 \sin^2\theta)^{\frac{\eta-\xi}{2}}}{(R^2 + a^2 \cos^2\theta)^{\frac{\eta-\xi+1}{2}}}. \tag{53}$$

Under the assumption that $a^2/R^2 \ll 1$, and assuming t=0 when θ=0, we have

$$t \cong \frac{R}{\left(1-\frac{2r_0}{R}\right)^{\frac{\eta-\xi}{2}}}\theta + \frac{a^2}{2R} \cdot \frac{\left[\left(1-\frac{2r_0}{R}\right)\left(\frac{\eta-\xi+1}{2}\right)+\frac{\xi-\eta}{2}\right]}{\left(1-\frac{2r_0}{R}\right)^{\frac{\eta-\xi+2}{2}}} \int_o^\theta \cos\theta' d\theta'$$

$$= \frac{R}{\left(1-\frac{2r_0}{R}\right)^{\frac{\eta-\xi}{2}}}\theta + \frac{a^2}{4R} \cdot \frac{\left[\left(1-\frac{2r_0}{R}\right)\left(\frac{\eta-\xi+1}{2}\right)+\frac{\xi-\eta}{2}\right]}{\left(1-\frac{2r_0}{R}\right)^{\frac{\eta-\xi+2}{2}}} (\cos\theta \sin\theta + \theta)$$

$$= \left(\frac{R}{\left(1-\frac{2r_0}{R}\right)^{\frac{\eta-\xi}{2}}} + \frac{a^2}{4R} \cdot \frac{\left[\left(1-\frac{2r_0}{R}\right)\left(\frac{\eta-\xi+1}{2}\right)+\frac{\xi-\eta}{2}\right]}{\left(1-\frac{2r_0}{R}\right)^{\frac{\eta-\xi+2}{2}}}\right)\theta +$$

$$\frac{a^2}{8R} \cdot \frac{\left[\left(1-\frac{2r_0}{R}\right)\left(\frac{\eta-\xi+1}{2}\right)+\frac{\xi-\eta}{2}\right]}{\left(1-\frac{2r_0}{R}\right)^{\frac{\eta-\xi+2}{2}}} \sin 2\theta. \tag{54}$$

During this time, the rotating observer describes an angle $\theta_0$ while light travels an angle $2\pi \pm \theta_0$ (once again, + for co-rotating beam and - for the counterrotating beam) so that

$$\frac{\theta_0}{\omega_0} = (p+q)(2\pi \pm \theta_0) \pm \frac{q}{2}\sin 2\theta_0 \tag{55}$$

where

$$q = \frac{a^2}{4R} \cdot \frac{\left[\left(1-\frac{2r_0}{R}\right)\left(\frac{\eta-\xi+1}{2}\right)+\frac{\xi-\eta}{2}\right]}{\left(1-\frac{2r_0}{R}\right)^{\frac{\eta-\xi+2}{2}}}, \quad p = \frac{R}{\left(1-\frac{2r_0}{R}\right)^{\frac{\eta-\xi}{2}}}. \tag{56}$$

Assume, as we did already, a low speed observer and that the angle $2\theta_0$ be so small as to justify $\sin 2\theta_0 \cong 2\theta_0$. Then

$$\frac{\theta_0}{\omega_0} = (p+q)(2\pi \pm \theta_0) \pm q\theta_0. \tag{57}$$

Solving for $\theta_0$, we get

$$\theta_{0\pm} = 2\pi \frac{p+q}{\dfrac{1}{\omega_0} \mp (p+q) \mp q}. \tag{58}$$

Finally, the difference between two round trip "coordinate" times (recalling the approximations already used) comes to

$$t_+ - t_- = \frac{\theta_{0+} - \theta_{0-}}{\omega_0}$$

$$= 4\pi\omega_0 \cdot \frac{\left[X + \dfrac{a^2}{2R}Y\right]\left[X + \dfrac{a^2}{R}Y\right]}{Z^2 - \omega_0^2\left[X + \dfrac{a^2}{R}Y\right]^2} \tag{59}$$

where

$$X = R(1 - \frac{2r_0}{R}) \tag{60}$$

$$Y = (1 - \frac{2r_0}{R})\left(\frac{\eta - \xi + 1}{2}\right) + \frac{\xi - \eta}{2} \tag{61}$$

$$Z = \left[1 - \frac{2r_0}{R}\right]^{\frac{\eta - \xi + 2}{2}} \tag{62}$$

Neglecting terms of order $R^{-3}$ and $\omega_0^2 R^2$ and higher, we get

$$t_+ - t_- \cong \pi\omega_0 R^2 \left\{4 + \frac{3a^2}{R^2} + \frac{8r_0(\eta - \xi)}{R}\right\}. \tag{63}$$

Thus, the correction due to the angular momentum of the source is independent of R in this case. The term is in fact given by, using Eq.(52),

$$3\pi a^2 \omega_0 = \frac{12}{25}\pi R_0^4 \Omega_0 \omega_0 \qquad (64)$$

where $R_0$ is the radius of a source sphere of uniform density.

In order to obtain what the rotating observer measures, we must calculate the proper time in his/her frame. This is done as follows: From the metric (9),

$$\tau = \int \left[\left(1 - \frac{2r_0 R}{\rho}\right)^\eta - \left(1 - \frac{2r_0 R}{\rho}\right)^\xi \rho \omega_0^2\right]^{\frac{1}{2}} dt, \quad \rho = R^2 + a^2 \cos^2(\omega_0 t). \qquad (65)$$

For short enough $\omega_0 t$, we have, $\cos(\omega_0 t) \cong 1$, $\sin(\omega_0 t) \cong 0$. Further, neglecting terms of the order $R^{-2}$ in the integrand, we have

$$\tau \cong \left(1 - \frac{2r_0 \eta}{R} - \omega_0^2 R^2\right)^{\frac{1}{2}} t. \qquad (66)$$

Therefore, the time delay in the polar case, denoted by $\delta\tau_P$, is given by

$$\delta\tau_P \cong \left(1 - \frac{2r_0 \eta}{R} - \omega_0^2 R^2\right)^{\frac{1}{2}} (t_+ - t_-)$$

$$= \left(1 - \frac{2r_0 \eta}{R} - \omega_0^2 R^2\right)^{\frac{1}{2}} \left[\pi \omega_0 R^2 \left\{4 + \frac{3a^2}{R^2} + \frac{8r_0(\eta - \xi)}{R}\right\}\right]. \qquad (67)$$

Therefore, to the first and second orders in $\zeta$, $\psi$ and $\varepsilon$, we have

$$\delta\tau_P \cong \delta\tau_S (1 - \eta\varepsilon)\left[1 + \frac{3}{4}\zeta^2 + 2(\eta - \xi)\right]$$

$$\cong \delta\tau_S \left[1 + \frac{3}{4}\zeta^2 + (\eta - 2\xi)\varepsilon\right].$$

Comparing with the equatorial case, the excess is, using Eq.(49),

$$\frac{\Delta\tau}{\delta\tau_S} = \frac{\delta\tau_E - \delta\tau_P}{\delta\tau_S} \cong \frac{2(\sigma - \eta)\zeta\varepsilon}{\psi} - \frac{3}{4}\zeta^2. \qquad (68)$$

The term $(\eta-2\xi)\varepsilon$ cancels out due to the spherical symmetry of the orbits considered. After cross multiplying by $\delta\tau_S$, we get

$$\Delta\tau \equiv \delta\tau_E - \delta\tau_p \cong \frac{8\pi r_0 a}{R}(\sigma - \eta) - 3a^2\pi\omega_0. \tag{69}$$

It may be observed from Eqs.(50) and (69) that the scalar field appears only in the terms that contain the gravitating mass parameter $r_0$. This fact is quite consistent with the form of the KB metric which also has this property.

## 6. GEODESICS

Let us now consider the geodesic motion of the source/receiver having a 4-velocity $u^\mu (\equiv dx^\mu/ds)$. The geodesic equations are

$$\frac{\partial u^\mu}{\partial x^\nu} u^\nu + \Gamma^\mu_{\nu\alpha} u^\nu u^\alpha = 0 \tag{70}$$

where $\Gamma^\mu_{\nu\alpha}$ are the Christoffel symbols formed from the KB metric (9). We can simplify the problem by taking $\theta=\pi/2$, that is, $u^\theta=0$. The geodesic equations do allow such a solution [12]. In this case, $\sin\theta=1$, $\cos\theta=0$, $\omega=a$ and $P=1-2r_0/r$. For a circular geodesic orbit with a constant radius $r=R$, the condition is $u^r=0$. Then the radial equation becomes

$$\Gamma^r_{tt}(u^t)^2 + \Gamma^r_{\varphi\varphi}(u^\varphi)^2 + 2\Gamma^r_{t\varphi}u^t u^\varphi = 0. \tag{71}$$

Defining the angular speed of rotation of the source/receiver as $\omega=u^\varphi/u^t$, we get

$$\omega_\pm = \frac{1}{\Gamma^r_{\varphi\varphi}}\left[-\Gamma^r_{t\varphi} \pm \sqrt{\left(\Gamma^r_{t\varphi}\right)^2 - \Gamma^r_{tt}\Gamma^r_{\varphi\varphi}}\right]. \tag{72}$$

The above expression simply turns out to be

$$\omega_\pm = \frac{1}{\frac{\partial g_{\varphi\varphi}}{\partial r}}\left[-\frac{\partial g_{t\varphi}}{\partial r} \pm \sqrt{\left(\frac{\partial g_{t\varphi}}{\partial r}\right)^2 - \frac{\partial g_{tt}}{\partial r}\cdot\frac{\partial g_{\varphi\varphi}}{\partial r}}\right] \tag{73}$$

where

$$\frac{\partial g_{\varphi\varphi}}{\partial r} = \left(\frac{2}{r^2}\right)\left[\eta a^2 r_0 P^{\eta-1} - r^3 P^{\xi} - \xi r^2 r_0 P^{\xi-1} - 2\sigma a^2 r_0 P^{\sigma-1}\right] \qquad (74)$$

$$\frac{\partial g_{t\varphi}}{\partial r} = -\left(\frac{2}{r^2}\right)\left[a\eta r_0 P^{\eta-1} - a\sigma r_0 P^{\sigma-1}\right] \qquad (75)$$

$$\frac{\partial g_{tt}}{\partial r} = \left(\frac{2}{r^2}\right)\left[\eta r_0 P^{\eta-1}\right] \qquad (76)$$

Thus, at r=R, we finally have, $P=1-2r_0/R$ and

$$\omega_{\pm} = \frac{\widetilde{P}}{\widetilde{Q}} \qquad (77)$$

where

$$\widetilde{P} \equiv aM\left(\eta P^{\eta-1} - \sigma P^{\sigma-1}\right)$$
$$\pm \left[\eta R^3 r_0 P^{\xi+\eta-1} + a^2\sigma^2 r_0^2 P^{2\sigma-2} + \eta\xi R^2 r_0^2 P^{\xi+\eta-2}\right]^{\frac{1}{2}}$$

$$\widetilde{Q} \equiv \eta a^2 r_0 P^{\eta-1} - R^3 P^{\xi} - \xi R^2 r_0 P^{\xi-1} - 2\sigma a^2 r_0 P^{\sigma-1}$$

Dividing the numerator and denominator of $\omega_{\pm}$ by $R^3 P^{\xi}$ and retaining terms up to a/R, we find

$$\omega_{\pm} \approx \mp \frac{1}{R}\sqrt{\frac{\eta r_0}{R}} + \frac{ar_0}{R^3}(\sigma - \eta). \qquad (78)$$

The sign flip in this equation can be rectified. Suppose we follow the convention that $\omega_+ > 0$ and $\omega_- < 0$ in the Kerr limit, that is, the $\pm$ signs on $\omega_{\pm}$ indicate the sign of the frequency. Then, from Eq.(75), assuming a>0, we find that $\partial g_{t\varphi}/\partial r < 0,$ so that the numerator (the big []) in Eq.(73) is positive. But $\partial g_{\varphi\varphi}/\partial\varphi$ in Eq.(74) has the leading term $-2rP^{\xi} < 0$. Thus $\omega_+$ as defined by eq. (73) is actually negative in the Kerr limit and similarly $\omega_- > 0$. Thus if we were to change the $\pm \rightarrow \mp$ on the right hand side of Eq.(73) (in the big []) then eq. (78) would read, using the notations of Sec. 4, as

$$\psi_\pm \equiv \omega_\pm R \cong \pm\sqrt{\eta\varepsilon} + (\sigma - \eta)\varepsilon\zeta. \qquad (79)$$

On using this in Eq.(49), we get the delay

$$\delta\tau_{G\pm} \cong 4\pi R[\psi_\pm + 2(\sigma - \eta)\varepsilon\zeta + (\eta - 2\xi)\varepsilon\psi_\pm]$$

which yields, to the lowest order in ε:

$$\delta\tau_{G\pm} \cong 4\pi R\left[\pm\sqrt{\eta\varepsilon} + 3(\sigma - \eta)\varepsilon\zeta + O(\varepsilon)^{\frac{3}{2}}\right]. \qquad (80)$$

Now the traditional Sagnac effect is [12], obtained here by putting in Eq.(80), a=0, η=1 and σ=0:

$$\delta\tau_{S\pm} = 4\pi R\psi_\pm = \pm 4\pi\sqrt{MR}$$

so that we have

$$\delta\tau_{G\pm} \cong \sqrt{\frac{\eta r_0}{M}}\delta\tau_{S\pm} + \frac{12\pi a r_0}{R}(\sigma - \eta) + O\left(\frac{r_0}{R}\right)^{\frac{3}{2}}. \qquad (81)$$

Thus, unlike the case of polar or equatorial orbits, the traditional part of the Sagnac effect is multiplied by a factor $\sqrt{\frac{\eta r_0}{M}}$. Its value will be found from the PPN form of the metric (9) in Sec.8.

## 7. EINSTEIN FRAME

It is instructive to calculate the relevant corrections in the Einstein frame as well, already defined in Sec.2. The metric to be used now is (17) and the steps to be followed are precisely the same as those in Sections 3-6. However, it is not necessary to do them explicitly. Instead, one may simply use the replacements given by η→η-σ, ξ→ξ-σ and σ→σ-σ in the desired expressions computed in the Jordan frame.

(a) Equatorial Orbits

As can be verified, $\omega_n$ of Eq.(33) remains completely unaffected, that is, $\omega_n^{(J)} = \omega_n^{(E)}$. This implies that the definition of "static" observers, for which no Sagnac delay exists, is preserved even though the physics in the two frames differ widely. However, $\delta\tau_0$ of Eq.(39) changes to

$$\delta\tau_0^{(E)} = \delta\tau_0^{(J)} \approx 8\pi\omega_{LT} R^2 (\sigma - \eta) \tag{82}$$

The exact expression for the delay, that is, $\delta\tau$ between the two frames are also related in the same way and under the approximations as before, we find, from Eq.(50):

$$\delta\tau^{(E)} = \delta\tau^{(J)} \cong \delta\tau_S + \frac{8\pi r_0 a}{R}(\sigma - \eta) + 4\pi r_0 R \omega_0 (\eta - 2\xi + \sigma). \tag{83}$$

(b) Polar Orbits

It can easily be noticed from the eqs.(56) that $p^{(E)} = p^{(J)}, q^{(E)} = q^{(J)}$ so that we have $(t_+ - t_-)^{(E)} = (t_+ - t_-)^{(J)}$ and consequently, from Eq.(68):

$$\delta\tau_P^{(E)} \cong \delta\tau_S \left[1 + \frac{3}{4}\varsigma^2 + (\eta - 2\xi + \sigma)\varepsilon\right]. \tag{84}$$

The difference becomes, using Eq.(69),

$$\Delta\tau^{(E)} = (\delta\tau_E - \delta\tau_P)^{(E)}$$

$$= \frac{8\pi a r_0}{R}(\sigma - \eta) - 3\pi a^2 \omega_0. \tag{85}$$

(c) Geodesics

The exact expression for $\omega_\pm^{(E)}$ can be easily obtained from Eq.(77) under the specified replacements. We shall here write only the approximated final result, from Eq.(81):

$$\delta\tau_{G\pm}^{(E)} = \sqrt{\frac{(\eta - \sigma)r_0}{M}}\delta\tau_{S\pm} + \frac{12 r_0 a(\sigma - \eta)}{R} + O\left(\frac{r_0}{R}\right)^{\frac{3}{2}}. \tag{86}$$

Although some of the terms in eqs.(83), (84) and (86) look different from the corresponding terms in the Jordan frame, a PPN approximation will show that they are actually the same. In fact, the coefficients in the first terms in eqs.(81) and (86) are both unity!

## 8. DISCUSSIONS
### 8.1. STR Numerical Estimates

In the foregoing, we calculated the effect of the BD scalar field on the gravitational corrections to the Sagnac effect in the Jordan and Einstein frames. Three types of source/observer trajectories were considered, viz., equatorial, polar and geodesic. In the Jordan frame the corresponding expressions are eqs.(50), (69) and (81), while in the Einstein frame, these are eqs.(83), (85) and (86). All these expressions reveal the effect of the scalar field through the presence of $\eta$, $\xi$ and $\sigma$. Since these parameters are connected by Eq.(10), it is clear that the knowledge of any two would suffice in determining the remaining one. Measurements of the correction terms would place upper limits on the values of $\eta$ and $\sigma$. These limits would translate into a limit on $\varpi$, via eqs.(8) and (10), just as it happened in the static BD solutions with respect to solar system tests. Conversely, we can take the solar system value $\varpi \geq 500$ and calculate the expected numerical values of $\eta, \sigma$ and $\xi$.

For the sake of comparison, let us now estimate the numerical values of the basic as well as the correction terms in STR. Consider the exact proper time delay $\delta\tau$ from STR given by (under similar circumstances as in Sec.3):

$$\delta\tau_{STR} = \frac{(4\pi R^2)(\omega_0 + \Omega)}{\sqrt{(1 - \Omega^2 R^2) - 2\omega_0 \Omega R^2 - \omega_0^2 R^2}} \tag{87}$$

where $\Omega$ and $\omega_0$ are, respectively, the angular speed of the coordinate system rotating about the origin (turntable) and orbital angular speed of the source/observer with respect to this turntable [28]. If the coordinate system is nonrotating, that is $\Omega=0$ but $\omega_0 \neq 0$, then

$$\delta\tau_{\Omega=0} = (4\pi\omega_0 R^2)(1 - \omega_0^2 R^2)^{-\frac{1}{2}} \tag{88}$$

and conversely, if the source/observer is fixed to the turntable such that $\omega_0=0$ but $\Omega\neq 0$, then

$$\delta\tau_{\omega_0=0} = (4\pi\Omega R^2)(1-\Omega^2 R^2)^{-\frac{1}{2}}, \tag{89}$$

The effect is doubled if the source/observer has $\omega_0=\Omega\neq 0$

$$\delta\tau(\omega_0\equiv\Omega) = (8\pi\Omega R^2)(1-4\Omega^2 R^2)^{-\frac{1}{2}} \tag{90}$$

and is zero if $\omega_0=-\Omega$, that is, when the source observer is moving on the turntable opposite to its rotation but with the same angular speed $\Omega$.

Tartaglia [12] considers the case when the source/observer is fixed to the equator of the Earth, which means one has to consider Eq.(89) with $\Omega=\Omega_\oplus$ where the symbol $\oplus$ denotes Earth values. Expanding Eq.(89), and restoring c, we get

$$\delta\tau(\omega_0=0) = \frac{4\pi\Omega_\oplus R_\oplus^2}{c^2} + \frac{2\pi\Omega_\oplus^3 R_\oplus^4}{c^4} + ... \tag{91}$$

where $R_\oplus$ denotes the radius of the Earth.

Now recall the relevant data for Earth:

$$R_\oplus = 6.37\times 10^6 \text{ m}$$
$$\Omega_\oplus = 7.27\times 10^{-5} \text{ rad/s}$$
$$\frac{GM_\oplus}{c^2} = 4.4\times 10^{-3} \text{ m}$$
$$a_\oplus = 9.81\times 10^8 \text{ m}^2/\text{s}$$
$$c = 3\times 10^8 \text{ m/s}.$$

Substituting these values into Eq.(89) we obtain

$$\delta\tau_{STR}(\omega_0=0) = [4.12\times 10^{-7} + 4.6\times 10^{-19} + ...] \text{ s} \tag{92}$$

Therefore, the basic Sagnac delay Eq.(1), amounts to $4.12\times 10^{-7}$s. To compare the above terms with the corresponding ones in the BD theory, we must first determine the unknown constants appearing there. This is achieved by using the PPN approximation, discussed below.

8.2. PPN Approximation

Our aim in this subsection is to express the KB parameters $\eta,\sigma,\xi$ in terms of the coupling constant $\varpi$. The first step in this direction is to rewrite our eq.(8) in the form

$$1-(\eta-\sigma)^2 = (2\varpi+3)\sigma^2 \qquad (93)$$

by noting that

$$\eta = \frac{1}{\lambda}, \sigma = -\frac{C}{2\lambda}, \xi = \frac{\lambda-C-1}{\lambda}. \qquad (94)$$

The next step is to consider the PPN parameters $\alpha,\beta,\gamma$ which appear in the metric

$$ds^2 \approx -\left[1-2\alpha\left(\frac{M}{\rho}\right)+2\beta\left(\frac{M}{\rho}\right)^2\right]dt^2 + \left[1+2\gamma\left(\frac{M}{\rho}\right)\right](d\rho^2+\rho^2 d\Omega^2) \qquad (95)$$

Since $\eta,\sigma,\xi$ already appear in the static form of the metric (9), and we are considering only the weak field form of the metric, we can, for the moment, assume a=0. In isotropic coordinates $(\rho,\theta,\varphi)$ given by

$$r = \rho\left(1+\frac{r_0}{2\rho}\right)^2,$$

the reduced metric (9) becomes

$$ds^2 = -\left[\frac{1-\frac{r_0}{2\rho}}{1+\frac{r_0}{2\rho}}\right]^{2\eta} dt^2 + \left[\frac{1-\frac{r_0}{2\rho}}{1+\frac{r_0}{2\rho}}\right]^{2\xi-2}\left(1+\frac{r_0}{2\rho}\right)^4 (d\rho^2+\rho^2 d\Omega^2). \qquad (96)$$

Comparing the corresponding orders, we get

$$\alpha = 1, \beta = 1, \gamma = 1-\frac{2\sigma}{\eta}, \eta r_0 = M. \qquad (97)$$

The usual PPN value of $\gamma$ is $\gamma=(1+\varpi)/(2+\varpi)$ [29] and using eq.(92), we get

$$\sigma = \frac{1}{\sqrt{(2\varpi+3)(2\varpi+4)}}, \eta = \sqrt{\frac{2\varpi+4}{2\varpi+3}}, \xi = 1-\eta+2\sigma. \qquad (98)$$

Let us now consider the weak field rotational part given by $4(\eta-\sigma)\dfrac{r_0}{\rho^3}(xdy-ydx)dt$ (See later, in Sec.9). Using $r_0 = M/\eta$, we find that the effect of the scalar field is equivalent to multiplying the Kerr part by the factor $\left(\dfrac{2\varpi+3}{2\varpi+4}\right)$, which is exactly the PPN prediction as well.

Regarding the values given in eqs.(98) as those determined from the weak field boundary conditions, we can now rewrite the exact form of Sagnac delay given in eq.(35):

$$|\delta\tau_0| = (4\pi a) \times \left[\frac{\left(1-\dfrac{2M}{\eta R}\right)^\eta - \left(1-\dfrac{2M}{\eta R}\right)^\sigma}{\left(1-\dfrac{2M}{\eta R}\right)^{\tfrac{\eta}{2}}}\right] \qquad (99)$$

which yields, to second order in $(M/R)^2$:

$$|\delta\tau_0| \approx \frac{8\pi aM}{R}\left[\frac{2\varpi+3}{2\varpi+4}\right]\left[1+\frac{M}{R}\left(\frac{1}{2\varpi+4}-\sqrt{\frac{2\varpi+3}{2\varpi+4}}\right)\right] \qquad (100)$$

Eq.(50) represents the corrections due to other physical factors (such as the moment of inertia etc) and using the boundary values in eqs.(98), one can easily deduce how the scalar field combines with them through the appearance (or absence) of $\varpi$.

An exact expression for the Sagnac delay for polar orbits can be obtained by plugging in the value of $(t_+ - t_-)$ from eq.(59) into eq.(65). A similar expression can be obtained for the geodesic motion using eqs.(45), (77) and (79). Expansion of these exact expressions would enable us to assess the influence of other physical factors as well as the involvement of the scalar field.

A simple demonstration will reveal that calculations in both the Jordan and Einstein frames lead to the *same* $\varpi$ factors for the corrections. Turning to the calculations in the Einstein frame for which the KB metric is given by eq.(18), we find from the PPN requirement that

$$(\eta-\sigma)r_0 = M, \sigma-\eta \to \sigma-\eta = \sqrt{\frac{2\varpi+3}{2\varpi+4}},\ \eta \to \eta-\sigma = \sqrt{\frac{2\varpi+4}{2\varpi+3}}, \qquad (101)$$

$$1-(\eta-\sigma)^2 = (2\varpi+3)\sigma^2.$$

Then the first order correction term in eq.(83) reads

$$\frac{8\pi aM}{R}\left[\frac{2\varpi+3}{2\varpi+4}\right]$$

which is precisely the same as the first term in eq.(100). Use of eqs.(101) would enable us to see also that eqs. (50) and (83), (69) and (85), (81) and (86) are actually the same.

8.3. BD Numerical Estimates

In order to compare Eq.(92) with the corresponding situation in the BD theory, we should consider the case when the source/observer is fixed on the surface of the Earth, viz., $\omega_0=\Omega_\oplus$. The various correction terms are, for the equatorial orbit, putting $\omega_0=\Omega_\oplus$ in Eq.(50) and using the identification $r_0 = M/\eta$:

$$\frac{4\pi GM_\oplus R_\oplus \Omega_\oplus}{c^4}(1-\frac{2\xi}{\eta}) \approx 2.84\times 10^{-16}(1-\frac{2\xi}{\eta})\text{ s} \qquad (102)$$

$$\frac{8\pi GM_\oplus a_\oplus}{R_\oplus c^4}(\frac{\sigma}{\eta}-1) \approx 1.89\times 10^{-16}(\frac{\sigma}{\eta}-1)\text{ s} \qquad (103)$$

These estimates suggest that the first two corrections in Eq.(50) are at least three orders of magnitude higher than the STR one if $\eta$ and $\sigma$ assume nearly Kerr values. For visible light, $\nu \sim 10^{14}$Hz, and ignoring for the moment the BD parameters $(1-2\xi/\eta)$ and $(\sigma/\eta-1)$, the expected fringe shift would be $\sim 10^{-2}$ and the parameters would alter the above multiplicative coefficients. Thus, depending on the deviation of the observed shift from this resulting value, we might conclude about the existence of BD scalar field.

In computing the polar and geodesic cases, Tartaglia [12] considers polar and geodesic trajectories of the same radius $R = 7\times 10^6$ m. Then, our Eq.(68) for polar orbits reveals the following: If we take $\omega_0 = \frac{1}{R}\sqrt{\frac{GM}{R}}$, the first and the second terms are of order $\sim 10^{-15}$ $(1-2\xi/\eta)$ s and $\sim 10^{-18}$ s respectively. Considering the first term, one has an expected fringe shift of order $\sim 10^{-1}(1-2\xi/\eta)$ s for visible light. From the difference in Eq.(69), we find that the first term on the r.h.s. is of $\sim 10^{-16}$ $(\sigma/\eta-1)$ s, or equivalent to $10^{-2}$ $(\sigma/\eta-1)$ fringe shift but the advantage of this equation is that one need not fix a "zero" or a "pure" Sagnac term (that is, the one unaffected by either gravity or scalar field).

For a circularly orbiting geodesic source/observer (Earth bound satellites, for example) with an orbit radius, say, $R=7\times10^6$ m, the first term on the r.h.s. of Eq.(81) is $7.35\times10^{-6}$s. This delay corresponds to a fringe shift of $\sim10^8$ for visible light, which should be immensely measurable. A first order correction to this, namely, the second term in eq.(81) is of the order $\sim10^{-16}$ ($\sigma/\eta$-1)s. Therefore, a better correction term still follows from eqs.(50) [which is of the order of $\sim10^{-1}(1-2\xi/\eta)$ ] and it would put bounds on $\varpi$. One then has to compare these bounds with the Kerr values in order to determine whether a BD scalar field is feasible or not. Even if we take the lowest value for $\varpi$, viz., $\varpi= 500$, the coefficients in eqs.(102) and (103) respectively would change only very minutely. Accordingly, the required measurement has to be very precise so that such small deviations are detectable. Feasibilities of such measurements are discussed next.

## 8.4. Optical and matter-wave interferometric measurements

Bounds on $\varpi$ at least from the leading term $\frac{8\pi aM}{R}\left(\frac{2\varpi+3}{2\varpi+4}\right)$ should be within the realm of experimental feasibility. The discussion in the previous section reveals that Earth-bound verification of the Kerr and/or BD corrections to the basic Sagnac effect requires the detection of delays $O(10^{-14}-10^{-18}s)$ or $O(1-10^{-4})$ fringes, or equivalently $O(10^{-6}-10^{-10})\Omega_\oplus$ in interferometry experiments. In single-input-port optical gyroscopes and rotation sensors the minimal detectable phase scales as $\Delta\phi = O(1/\sqrt{N})$, where N is the number of particles passing through the device per unit time [30]. Currently devices are operating near this shot noise limit and can detect angular velocities of $O(10^{-10})\Omega_\oplus$ [31].

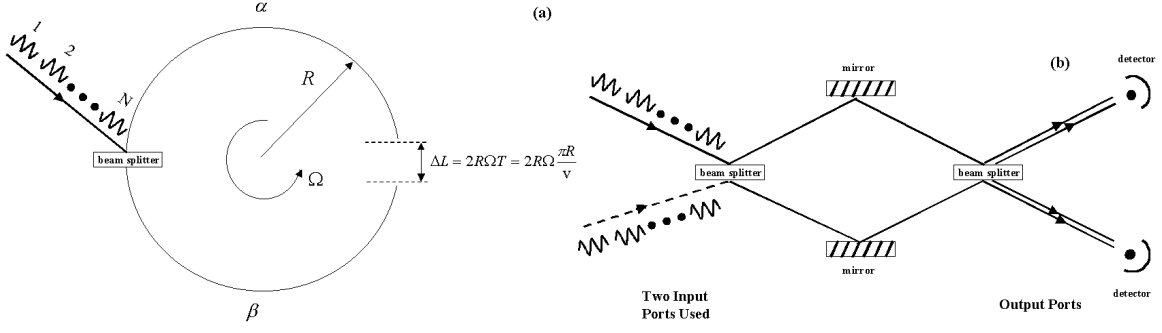

**Fig.(1)** (a): A schematic illustration of an idealized light or matter-wave interferometer used as a rotation sensor or gyroscope (after [30]). The interferometer has circular arms of radius R and rotates with angular velocity $\Omega$, with N atoms passing one-at-a-time through a beam splitter. The path difference between the upper and lower branches $\alpha$ and $\beta$ is given by $\Delta L = 2R\Omega T$ where $T = 2\pi R/c$ for light and $T = \pi R/v$ for matter. (b): A two input-port quantum interferometer. Quantum states are entangled (correlated) at the input ports and phase shifts are measured at the output ports. The use of correlated quantum states in the interferometer allows for minimum phase sensitivities which scale as $\Delta\phi = O(1/N)$ versus the uncorrelated state shot-noise limit of $\Delta\phi = O(1/\sqrt{N})$.

On the other hand, the use of material particles instead of light holds great promise in the field of interferometry and rotational sensors. The advantage of using matter over light in interferometers can be seen as follows: consider interferometer with semicircular arms rotating with angular frequency $\Omega$ about an axis through its center and normal to the loop plane depicted in Fig.(1a). In a given time T, particles traversing in the same and opposite rotational sense as the interferometer will travel a distance $L_+ = 2\pi R + R\Omega T$ and $L_- = 2\pi R - R\Omega T$ respectively, yielding a path difference of $\Delta L = 2R\Omega T$. For light with a single beam splitter input/output port we have $T = 2\pi R/c$, so that we recover Eq.(1) via $\delta\tau_S = \Delta L/c$. However, for particles of mass m traveling at velocity v, with a beam splitter output port located diametrically opposite the input port we have $T = \pi R/v$. This leads to $\delta\phi_{matter} = k\Delta L = 2A\Omega/\lambda_r v$ where $\lambda_r = \lambda/2\pi$ is the reduced wavelength. For matter, $\lambda_r = \hbar/mv$ is the de Broglie wavelength and the phase signal is given by $\delta\phi_{matter} = 2A\Omega m/\hbar$. For light, we can define the "photon mass" by $m_\gamma c^2 = \hbar\omega$. Thus the inherent sensitivity of a matter-wave interferometer exceeds that of a photon-based system by the mass-enhancement factor $mc^2/\hbar\omega \approx 10^{10-11}$. This impressive mass-enhancement factor for matter-wave interferometers is offset by a factor of $O(10^4)$ for smaller particle fluxes and $O(10^4)$ smaller number of cavity round trips (usually 1 for matter and $10^4$ for light). Matter-wave interferometry experiments have seen to date a sensitivity of $2 \times 10^{-8} (rad/s)\sqrt{Hz}$ [32], which is comparable to the best active ring laser gyroscopes, and they are getting better.

The use of quantum entangled input states or correlated-two-input-port interferometers offers exciting possibilities for the future [30]. A single input-port interferometer can be considered as a two-input-port device where light or matter enters in one port (i.e. one side of a beam splitter) as the source and the ever present vacuum enters the second (empty) port. The minimal detectable phase scales as $\Delta\phi = O(1/\sqrt{N})$ where N is the number of particles passing through the device in unit time. In a two-input-port device, a non-vacuum state is presented to each port and is correlated at the input beam splitter as shown in Fig.(1b). The use of quantum entangled states (for both matter and light) leads to a minimal detectable phase sensitivity scales as $\Delta\phi = O(1/N)$. It can be shown that a two-input-port matter-wave interferometer can be $10^6$ more sensitive than a single-input-port matter-wave interferometer, a two-input-port optical interferometer can be $10^8$ times more sensitive than a single-port optical interferometer, and a two-input-port matter-wave interferometer can be an impressive $10^{10}$ times more sensitive than a single-input port optical interferometer.

Clearly there are considerable technical challenges to overcome in bringing such devices to fruition. *Decoherence*, the intrinsic quantum decay that ensues when a quantum system is coupled to undesired states, can degrade the performance of matter-wave or entangled quantum detectors and reduce the phase sensitivity back down to $\Delta\phi = O(1/\sqrt{N})$ [33]. This result can sometimes occur since, although the phase sensitivity increases with the number of particles N used in the interferometer, the decoherence rate grows commensurately. However, even with decoherence issues considered, current experiments are already making significant strides towards realizations of matter-wave and entangled quantum state interferometers useful for measuring the Sagnac effect [32] . With such promise, we may someday soon be able to experimentally detect the higher order general relativistic corrections to the Sagnac effect and be able to place tighter bounds on the BD parameters.

## 9. Geodetic and Lense-Thirring Precession

We can also investigate the effects of the KB metric on the precession of a spherical gyroscope in a circular polar orbit around the Earth as a means to experimentally measure or bound the values of the parameters $\eta, \xi, \sigma$ or just $\varpi$. The Stanford Gravity Probe-B Experiment [34] is just such an experiment which will use a superconducting niobium coated quartz spherical gyroscope (machined to a precision greater than $10^{-6}$ cm) to detect gravitational precession effects arising from the geodetic motion of the satellite and due to the rotation of the Earth (Lense-Thirring effect). In the following, we follow the calculation of Ohanian and Ruffini [35] by writing the KB metric to first order in $\varepsilon = r_0/r$ and $\zeta = a/r$, converting to isotropic coordinates and then computing the parallel transport equation for the spin $S^\mu$ of the gyroscope as it is carried about the polar circular orbit. Isotropic coordinates $(x, y, z)$ are used since a change in the rectangular components of the spin vector can be immediately attributed to the curvature of spacetime, whereas a change in curvilinear components contains contributions both from the curvature of the coordinates and the curvature of spacetime.

We begin with the KB metric in the Jordan frame, Eq.(9), and expand it to first order in $\varepsilon, \zeta$ to obtain

$$ds^2 \cong (1 - 2\eta\varepsilon)dt^2 - (1 - 2(\xi-1)\varepsilon)dr^2 - (1 - 2\xi\varepsilon)(r^2 d\theta^2 + r^2 \sin^2\theta d\phi^2) \\ + 4(\eta - \sigma)\varepsilon\zeta \sin^2\theta\, r d\phi\, dt \quad (104)$$

The change to a radial isotropic coordinate is the same as in the Schwarzschild case (see [36], p196ff and p256ff) and is given by $r = \rho(1 + r_0/2\rho)^2 \approx \rho(1 + r_0/\rho)$, where $\rho$ is the radial isotropic marker. To lowest order $\varepsilon \to \varepsilon' \equiv r_0/\rho$, $\zeta \to \zeta' = a/\rho$, and from now on we drop the primes on $\varepsilon, \zeta$. Carrying out the change to a radial isotropic coordinate and using coordinates $x = \rho \sin\theta \cos\phi$, $y = \rho \sin\theta \sin\phi$, $z = \rho \cos\theta$, $|d\vec{\rho}|^2 = dx^2 + dy^2 + dz^2$ fixed to the center of the Earth and non-rotating with respect to the distant stars, and noting $\rho^2 \sin^2\theta\, d\phi = x\, dy - y\, dx$, we arrive at

$$ds^2 \cong \left(1 - \frac{2\eta r_0}{\rho}\right)dt^2 - \left(1 + 2(1 - 2\xi)\frac{r_0}{\rho}\right)|d\vec{\rho}|^2 + 4(\eta - \sigma)\frac{r_0 a}{\rho^3}(x\, dy - y\, dx)dt \,. (105)$$

Comparison with the Kerr metric [35,36] allows us to identify the last term of Eq.(105) with the rotation of the mass M (where $r_0 = GM/c^2$). In going from the Kerr to the KB metric we have the identification $a_{KB} = (1 - \sigma/\eta) a_{Kerr}$, where a=-J/Mc is the angular momentum per unit mass of the rotating body (for a body rotating in the positive sense J>0, $a$ is negative, see [36] p258).

We are now interested in computing the change in the spatial components of the spin $S^\mu$ of a gyroscope in a circular polar orbit, as depicted in Fig. (2). We will first evaluate the parallel transport equations for the spin at a single point $\vec{\rho} = (0, \rho, 0)$ of the orbit where the 4-velocity is given by $\dot{x}^\mu \equiv dx^\mu/d\tau = (1, 0, 0, v)$ and where the velocity v of the satellite has a value on the order of $\sqrt{GM/\rho}$. The equation for the parallel transport of the spin is given by

$$\dot{S}^\mu \equiv \frac{dS^\mu}{d\tau} = -\Gamma^\mu_{\alpha\beta} S^\alpha \dot{x}^\beta . \qquad (106)$$

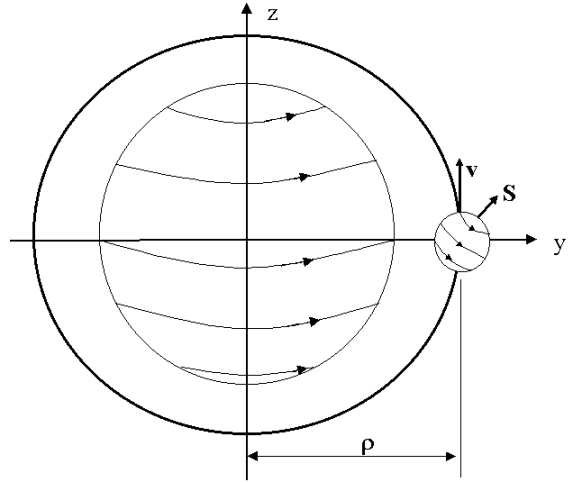

**Fig. (2)** A spherical gyroscope in a circular polar orbit about the Earth. At one instant, the gyroscope is at the position $x = 0, y = \rho, z = 0$ with instantaneous spatial velocity v along the $\hat{z}$ direction.

A lengthy, though straightforward, calculation yields the Christoffel symbols evaluated at the point $\vec{\rho} = (0, \rho, 0)$ to be

$$\Gamma^0_{02} = \eta r_0 / \rho^2, \Gamma^0_{12} = -3(\eta - \sigma) r_0 a / \rho^3,$$
$$\Gamma^1_{02} = -(\eta - \sigma) r_0 a / \rho^3, \Gamma^1_{12} = -(1 - 2\xi) r_0 a / \rho^2, \qquad (107)$$

$$\Gamma^2_{00} = \eta r_0/\rho^2, \Gamma^2_{01} = (\eta-\sigma)r_0 a/\rho^3, \Gamma^2_{11} = -\Gamma^2_{22} = \Gamma^2_{33} = (1-2\xi)r_0/\rho^2,$$
$$\Gamma^3_{23} = -(1-2\xi)r_0/\rho^2.$$

We note that in the co-moving reference frame of the satellite the spin is purely spatial $S'^0 = 0$, and the 4-velocity is purely temporal, $\dot{x}'^u = (1,0,0,0)$ so that the relationship $g'_{\alpha\beta}S'^\alpha \dot{x}'^\beta = 0$ holds. Since this is a tensor equation, it must also hold in the reference frame centered on the Earth, $g_{\alpha\beta}S^\alpha \dot{x}^\beta = 0$. This constraint allows us to solve for
$S^0 = -1/(g_{00} + v g_{03})\sum_{i=1,3}(g_{i0} + v g_{i3})S^i \cong v S^3 + O(\varepsilon)$. Substituting this and the Christoffel symbols into Eq.(106) yields the equations

$$\dot{S}^1 = \frac{r_0 a}{\rho^3} S^2,$$
$$\dot{S}^2 = -2\frac{r_0 v}{\rho^2} S^3 - \frac{r_0 a}{\rho^3} S^1, \qquad (108)$$
$$\dot{S}^3 = \frac{r_0 v}{\rho^2} S^2.$$

The terms proportional to v give rise to the geodetic precession while those proportional to *a* give rise to the Lense-Thirring precession. Although Eq.(108) was derived for a specific point on the orbit, we can generalize to any point on the orbit as follows. For *a*=0 we can write Eq.(108) as

$$\dot{\vec{S}}_g = -(1+\eta-2\xi)(\vec{v}\cdot\vec{S}_g)\vec{\nabla}\Phi + (1-2\xi)(\vec{S}_g \cdot \vec{\nabla}\Phi)\vec{v} \qquad (109)$$

where $\vec{S}_g$ refers to the geodetic contribution to the spin and $\Phi = -GM/\rho$ is the Newtonian gravitational potential. We are interested in the long-term secular change in the spin. As such we express the orbit of the satellite as $\vec{\rho} = \rho(0, \cos\omega_s t, \sin\omega_s t)$ where $\omega_s$ is the angular velocity of the satellite. Inserting $\vec{v} = d\vec{\rho}/dt = v(0, -\sin\omega_s t, \cos\omega_s t)$ where $v = \rho\omega_s$ and $\vec{\nabla}\Phi = r_0/\rho^2(0,\cos\omega_s t, \sin\omega_s t)$ into Eq.(109) and averaging over one period yields

$$\left\langle \vec{\dot{S}}_g \right\rangle = (1+\eta/2-2\xi)\frac{r_0 v}{\rho^2}\left(-S^3\,\hat{\mathbf{x}} + S^2\,\hat{\mathbf{y}}\right) \equiv \vec{\Omega}_g^{KB} \times \vec{S}_g$$

$$\vec{\Omega}_g^{KB} \equiv (1+\eta/2-2\xi)\frac{r_0}{\rho^3}\vec{\rho}\times\vec{v} = \frac{2}{3}(3/2-2\xi/\eta)\vec{\Omega}_g \qquad (110)$$

where $\vec{\Omega}_g^{KB}$ is the geodetic precession which reduces to the Schwarzschild and Kerr form $\vec{\Omega}_g = 3M/2\rho^3\,\vec{\rho}\times\vec{v}$ [35] in the limit $\{\eta\to 1, \xi=\sigma\to 0\}$. The geodetic precession of the spin $\vec{\Omega}_g$ is in the plane of the orbit and in the direction of the orbital motion of the satellite.

A similar calculation can be performed for the "gravitomagnetic" terms proportional to *a* in Eq.(108). These Lense-Thirring terms lead to precession of the spin in the direction perpendicular to the orbit and in the same sense as the rotation of the Earth ("frame dragging").

$$\vec{\Omega}_{LT}^{KB} = \frac{(\eta-\sigma)ar_0}{\rho^3}\left(\frac{3}{2}\left\langle(\vec{\rho}\cdot\hat{S}_\oplus)\vec{\rho}\right\rangle - \hat{S}_\oplus\right) = (1-\sigma/\eta)\vec{\Omega}_{LT}, \qquad (111)$$

where $\hat{S}_\oplus$ is a unit vector in the direction of the spin of the Earth (here $\hat{S}_\oplus = \hat{\mathbf{z}}$). As we observed earlier from the metric Eq.(105), this is just the usual Kerr Lense-Thirring precession $\vec{\Omega}_{LT}$ [35] with $a_{KB} = (1-\sigma/\eta)a_{Kerr}$. Performing the time average as above one obtains

$$\left\langle\vec{\Omega}_{LT}^{KB}\right\rangle = \frac{(\eta-\sigma)ar_0}{2\rho^3}\hat{S}_\oplus = (1-\sigma/\eta)\left\langle\vec{\Omega}_{LT}\right\rangle. \qquad (112)$$

For a 650-km circular polar orbit, as depicted in Fig.(2), with the spin of the satellite in the plane of the orbit, $v=\sqrt{GM/\rho}$ and $\{\eta\to 1,\xi=\sigma\to 0\}$ we obtain the values $|\vec{\Omega}_g| = 6.6''/yr, |\vec{\Omega}_{LT}| = 0.042''/yr$ [35]. Thus, for the KB metric in both the frames, these values would be multiplied by $2/3(3/2-2\xi/\eta)$ and $(1-\sigma/\eta)$ respectively [obtained by using $r_0 = M/\eta$ or $r_0 = M/(\eta-\sigma)$]. Since the Gravity Probe-B experiment is capable of measuring the bare $\{\eta\to 1,\xi=\sigma\to 0\}$ values of these precessions, any possible deviations due to the Kerr-like BD scalar field should be detectable.

## 10. SUMMARY

In the foregoing, our aim was to examine how the presence of a BD scalar field modifies the gravitational correction terms to the Sagnac effect. To our knowledge, such an analysis has not been undertaken heretofore. A first order effect on the geodetic and Lense-Thirring precession was also computed. It was found that the presence of the scalar field introduces a combination of different BD factors η, σ, ξ into the correction terms. The obtained results are of both theoretical and practical importance: The values of η and σ away from the Kerr values would indicate the presence of the BD scalar field.

The paper derives *exact* expressions for the scalar field modified Sagnac delay. The unknown BD factors can be determined in terms of $\varpi$ by using an input from the PPN analysis, viz., $\gamma = \dfrac{1+\varpi}{2+\varpi}$, as a *boundary* condition. From the expansion of the exact expressions, it is possible to directly find out corrections to *all* orders, visualize the physical characters of these terms and assess how the scalar field modifies each of them. Thus, the present formulation offers two distinct theoretical advantages: (1) It is applicable also in the strong field where the usual PPN analysis fails. (2) It has a flexibility in the sense that *any* functional choice of γ($\varpi$) is admissible leading to forms of η($\varpi$) and σ($\varpi$) different from those in eqs.(98). The possibility of a non-PPN γ and its physical implications are discussed in Ref.[39], but are not pursued in this paper.

From a practical standpoint, a first order fringe shift of $\sim 10^{-1}(1-2\xi/\eta)$ is predicted for the Sagnac delay for Earth bound equatorial orbits (R=7×10$^6$ m), which should be measurable given the accuracy being attained by the current technology. The most exciting promise is offered by the Stanford Gravity Probe-B experiment which is attempting to measure the geodetic and Lense-Thirring precessions for Earth bound orbits. As shown above, the mutiplying factors to the first order corrections are, respectively, $\dfrac{2}{3}\left(\dfrac{3}{2} - \dfrac{2\xi}{\eta}\right)$ and $1 - \dfrac{\sigma}{\eta}$. For an estimate, taking $\varpi$=500, we find, using the PPN values in eq.(98), that $\left|\dfrac{2\xi}{\eta}\right| \approx 2.98\times 10^{-3}, \left|\dfrac{\sigma}{\eta}\right| \approx 9.96\times 10^{-4}$.

It was demonstrated that the observable predictions in the two frames are identical, as expected. All the equations presented in this work reduce to those in the Kerr case. Lastly, Eq.(35) represents the exact BD expression for the gravitational analog of the Aharonov-Bohm effect [10,37,38]. We have to say more about this in a forthcoming paper.

## ACKNOWLEDGMENTS
One of us (KKN) is grateful to Dr. Arunava Bhadra, Center for High Energy and Cosmic Ray Physics, University of North Bengal, for stimulating discussions and technical assistance.

## REFERENCES
[1] G. Sagnac, C. R. Acad. Sci. Paris **157**,708 (1913).
[2] G.E. Stedman, Rep. Prog. Phys. v**60**, 615 (1997).
[3] L.D. Landau and E.M. Lifschitz, *The Classical Theory of Fields, Vol.2* (Pergaman Press, Oxford, 1975).
[4]. J.M. Cohen and H.E. Moses, Phys.Rev.Lett.,**39**,1641(1977); D.W. Allan and M.A.Weiss, Science **228**, 69 (1985).
[5] J.M.Cohen and B.Mashhoon, Phys.Lett. **A181**, 353(1993); B. Mashhoon, Phys. Lett. **A173**, 347 (1993).
[6] A.H. Rostomyan and A.M. Rostomyan, Phys. Stat. Sol. (a) **126**, 29 (1991).
[7] A. Werner, J. Staudenmann and R. Colella, Phys. Rev. Lett. **42**, 1103 (1979).
[8] F. Riehle, Th. Kisters, A. Witte, J. Helmcke and Ch.J. Borde, Phys. Rev. Lett. **67**, 177 (1991).
[9] J. Anandan, Phys. Rev. **D24**, 338 (1981).
[10] J.J. Sakurai, Phys. Rev. **D21**, 2993 (1980).
[11] M. Dresden and C.N. Yang, Phys. Rev. **D20**, 1846 (1979).
[12] A. Tartaglia, Phys. Rev. **D58**, 064009 (1998).
[13] C.H. Brans and R.H. Dicke, Phys. Rev. **124**, 925 (1961).
[14] B. Green, J.M. Schwarz and E. Witten, *Superstring Theory* (Cambridge University Press, Cambridge, 1987).
[15] G. Magnano and L.M. Sokolowski, Phys. Rev. **D50**, 5039 (1994).
[16] D. La and P.J. Steinhardt, Phys. Rev. Lett. **62**, 376 (1989); A.M. Laycock and A.R. Liddle, Phys. Rev. **D49**, 1827 (1994).

[17] E.W. Kolb, D. Salopek and M.S. Turner, Phys. Rev. **D42**, 3925 (1990); P.J. Steinhardt and F.S. Accetta, Phys. Rev. Lett. **64**, 2740 (1990); A.R. Liddle and D. Wands, Phys. Lett. **B293**, 32 (1992).
[18] H. Feng and L. Liu, Chinese Phys. Lett. **16**, 394 (1999).
[19] K.K. Nandi, J. Evans and A. Islam, Phys. Rev. **D55**, 2497 (1997).
[20] K.K. Nandi, B. Bhattacharjee, S.M.K. Alam and J. Evans, Phys.Rev. **D57**, 823(1998); P.E. Bloomfield, Phys. Rev. **D59**, 088501 (1999); K.K. Nandi, Phys. Rev. **D59**, 088502 (1999).
[21] K.K. Nandi, B. Bhattacharjee and S.M.K. Alam, Gen. Relat. Grav. **30**, 1331 (1998).
[22] L.A. Anchordoqui, S.P. Bergliaffa and D.F. Torres, Phys. Rev. **D55**, 5226 (1997).
[23] S. Cotsakis, P. Leach and G. Flessas, Phys. Rev. **D49**, 6489 (1994).
[24] E.T. Newman and A.I. Janis, J. Math. Phys. **6**, 915 (1965).
[25] K.D. Krori and D.R. Bhattacharjee, J. Math. Phys. **23**, 637 (1982).
[26] H.A. Buchdahl, Phys. Rev. **115**, 1325 (1959).
[27] A.G. Agnese and M. La Camera, Phys.Rev. **D31**, 1280 (1985).
[28] Take the rotating coordinate system of STR:
$$ds^2 = -(1 - \Omega^2 r^2)dt^2 + 2\Omega r^2 d\varphi dt + dr^2 + r^2 d\varphi^2 + dz^2$$
rotating with uniform angular velocity $\Omega$ about the z-axis. Let the source/observer be orbiting in a circle of radius r=R with angular speed $\omega_0$ on a plane so that dz=0. Then follow the procedure of Sec.3 to arrive at Eq.(87).
[29] R.V. Wagoner and D. Kalligas, "Scalar-tensor theories and gravitational radiation,", in *Relativistic Gravitation and Gravitational Radiation*, eds. J.-A. Marck and J.-P. Lasota (Cambridge University Press, 1997), 433-435.
[30] J.P. Dowling, Phys. Rev. **A57**, 4736 (1998), M.O. Scully and J.P. Dowling, Phys. Rev. **A48**, 3186 (1993).
[31] W. Schleich and M.O. Scully, General Relativity and modern optics, in Les Houches 1982, *New Trends in Atomic Physics*, ed. G. Grynberg and R. Stora (North-Holland, Amsterdam, 1984), 995-1124; G. Stedman *et al* Phys. Rev. **A51**, 4944 (1995); I. Ciufolini and J.A. Wheeler, *Gravitation and Inertia*, (Princeton Univ. Press, NJ, 1995).
[32] A. Lenet, T.D. Hammond, E.T. Smith, M.S. Chapman, R.A. Rubenstein and D.E. Pritchard, Phys. Rev. Lett. **78**, 760 (1997); T.L. Gustavson, P. Bouyer and M.A. Kasevich, Phys. Rev. Lett. **78**, 2046 (1997).

[33] S.F. Huelga, C. Macchiavello, T. Pellizzari, A.K. Ekert, M.B. Plenio and J.I. Cirac, Phys. Rev. Lett. **79**, 3865 (1997).

[34] Information and technical references on the Gravity Probe-B experiment can be found at http://www.einstein/standford.edu. See also J.P. Turneaure, C.W. Everitt and B.W. Parkinson, et al. "The Gravity-Probe-B Relativity Gyroscope Experiment", in R. Ruffini, ed. *Proceedings of the Fourth Marcel Grossmann Meeting on Relativity* (Elsevier, Amsterdam, 1986).

[35] H.C. Ohanian and R. Ruffini, *Gravitation and Spacetime*, 2$^{nd}$ ed., (W.W. Norton & Co., New York, 1994).

[36] R. Adler, M. Bazin and M. Schiffer, *Introduction to General Relativity*, (McGraw Hill, New York, 1975).

[37] M.D. Semon, Found.Phys. **12**, 49 (1982).

[38] E.G. Harris, Am. J. Phys. **64**, 378 (1996). An exact treatment is given in: P.M. Alsing, Am. J. Phys. **66**, 779 (1998).

[39] T.Matsuda, Prog.Theo.Phys., **47**, 738 (1972).